\documentclass[preprint]{aastex}
\begin{document}
\title{High Spectral Resolution H$_{2}$ Measurements of Herbig-Haro Objects
38, 46/47, and 120}
\author{Richard D. Schwartz\altaffilmark{1}} 
\affil{Department of Physics and Astronomy, University of 
   Missouri-St. Louis, 8001 Natural Bridge Road, St. Louis,
   MO 63121; schwartz@umsl.edu}

\and

\author{Thomas P. Greene\altaffilmark{1}}
\affil{NASA's Ames Research Center, MS 245-6, Moffett Field, CA
   94035-1000;thomas.p.greene@nasa.gov}
\altaffiltext{1}{Visiting astronomer at the Infrared Telescope Facility, which is
operated by the University of Hawaii under Cooperative Agreement no. NCC 5-538
with the National Aeronautics and Space Administration, Office of Space Science,
Planetary Astronomy Program.}

\begin{abstract}
We report high spectral resolution (R $\simeq$ 20,000) measurements of
the H$_{2}$ 1-0 S(1) line in Herbig-Haro Objects 38, 46/47, and 120.
The long-slit spectra reveal complex velocity structure with
evidence for bow-shock structures as well as prompt entrainment and shock
heating of ambient molecular gas.  Individual knots within HH 38 show
distinct double peaked velocity structure, consistent with that expected 
from spatially unresolved bow shocks.
A portion of the HH 47A bow shock is resolved in our
measurements, and the kinematics of the H$_{2}$ trace closely that
found for H$\alpha$ emission.  The evidence indicates that the preshock
medium for HH 47A that formed in the wake of a previous ejection contains
molecular clumps.  The HH 46CjetA feature in the HH 46/47 counterflow is
suggestive of a bow shock emerging from near the base of the flow, with
H$_{2}$ emission arising from ambient cloud material excited in the
oblique flanks of the bow shock.  HH 46F
is situated downstream at the boundary between the ovoid cavity and
the ambient cloud material, and represents either entrainment by a stellar 
wind that fills the cavity or entrainment in the far wing of a
giant bow shock.   For HH 120,
the H$_{2}$ velocity results also corroborate those found from atomic
emission, and there is substantial evidence that the preshock medium
is inhomogeneous, producing clumps of H$_{2}$ emission on the wings of
a bowshock that has an apex at HH 120A.   
\end{abstract}
\keywords{ISM: jets and outflows --- ISM: Herbig-Haro objects} 

\section{Introduction}

The role played by Herbig-Haro (HH) objects and jets from young stellar objects
(YSOs) in driving the more widespread bipolar molecular flows observed in
CO has been the target of recent investigations \citep{das02,dav02}.
The use of high resolution
spectroscopic instruments in the near-infrared has advanced our
understanding of the role of molecular hydrogen emission as a possible
intermediate manifestation of the coupling between the atomic outflows
and the much cooler CO outflows.  Long slit Echelle observations and
Fabry-Perot imaging have revealed the dynamics and kinematics of a
number of Herbig-Haro jets and HH shock structures. \citet{dav00}  
used Echelle spectra to investigate the shock structures of
a number of HH objects, finding evidence for turbulent boundary
layers and entrainment of material to form massive bipolar molecular
outflows.  \citet{sch99} used near-IR Echelle spectroscopy
to investigate the dynamics and kinematics of HH 43.
\citet{mov00} have applied Fabry-Perot
techniques to reveal the structure and dynamics of the HH 7-11 system.
Observations of near-infrared H$_{2}$ emission have the
additional advantages of suffering less extinction than optical observations
in the dark cloud complexes where most of the HH flows are located,
and serving as a probe to the molecular bulk of the flow material,
thus permitting a more complete picture of outflow systems.

\section{Observations and Data Reduction}

The NASA 3.0-m Infrared Telescope equipped with CSHELL, the single-order
Cryogenic Echelle Spectrograph \citep{tok90,gre93}, was used 
on the nights of 1996 January 8 and 9 to obtain spatially-resolved spectra
of structures in HH 38, HH 46/47, and HH 120.  Spectra of the H$_{2}$ 1-0 S(1)
emission line were acquired with a 1\farcs0 (5 pixel) wide slit, yielding
a spectral resolution of about 21,000 (14 km s$^{-1}$).  The detector was
a 256$\times$256 pixel InSb array, and order sorting was accomplished with
the use of stock circular variable filters (CVFs).  The slit length was 
30\arcsec~with an image scale of about 0\farcs2 pixel$^{-1}$.

In order to locate slit positions, direct images of each HH object were first
obtained in the 1-0 S(1) line of H$_{2}$ with 30 s exposures on target
followed by 30 s sky exposures.  Owing to the small field of view 
and vignetting in the CVF system, the extended structure of HH 38 could not be
obtained with a single exposure. Consequently, three exposures, offset from
one another by about 10\arcsec, were required to reveal the extended structure
and to locate appropriate positions for slit placement.  Spectra at each
position involved single 400 s exposures on target followed by 400 s sky
exposures.

An early-type standard (HR 1666) was observed at eight positions along the
slit to assess telluric absorption and to measure the degree of interference
fringing caused by internal reflections in the substrate cavities of the
CVFs.  As discussed in \citet{sch99}, no telluric absorption was
found in the vicinity of the 1-0 H$_{2}$ emission line, and the fringe
amplitudes were found to be about 10\% of the signal in the dispersion
direction with a smaller variation along the slit as well. 
Full-slit lamp spectra were used to obtain field distortion corrections
through use of the GEOTRAN task in IRAF.  The data reduction followed the
procedure discussed in \citet{sch99}.  Measurements of seven
telluric absorption lines in the spectra of standard stars agreed to
within about 3 km s$^{-1}$ with the Solar Atlas \citep{kur84} values.

Approximate flux calibrations for the H$_{2}$ emission were obtained from
observations of the standard stars HR 1666 and HR 1855.  Based upon the
seeing profile of the standard stars, it is estimated that only about
70\% of the light from the stars entered the 1\arcsec~slit, so the H$_{2}$
fluxes have been scaled upward by 30\%.  Given the uncertainties due to
fringing, the use of point sources (standard stars) to calibrate diffuse
sources, and atmospheric extinction, we estimate that the H$_{2}$ fluxes
have uncertainties from 30\% to 50\%.
 
\section{Data Analysis}

To delineate the portions of each of the objects observed, greyscale
images have been formed from the restricted direct field of the CSHELL
spectrograph.  The images, obtained in the 1-0 S(1) line of H$_{2}$, are
displayed in the left panels of Figs. 1 (HH 38), Fig. 2 (HH 47A), Fig.3 
(HH46Cjet), Fig. 4 (HH 46F), and Fig. 5 (HH 120).  Next to each direct
image panel is a drawing that indicates the orientation of the image and
the direction to the exciting star.  The slit position and width is drawn
with dashed lines on the direct image, and labeled in arcsec to correlate
with the position-velocity (p-v) panels.  The H$_{2}$ flux levels for the
lowest contour and the linear contour interval fluxes are given in the
figure captions.
We emphasize
that the direct images were taken only for the purpose of identifying
features for spectroscopy. With generally poor signal to noise, the
images do not show the detailed morphologies apparent in other H$_{2}$
imaging studies, such as those of \citet{eis94} on HH 46/47, 
\citet{hod95} on HH 120, and \citet{sta00} on the HH 38/43 system.

Velocity measurements of the emission knots indicated in Figs. 1-5 were
extracted using a 1 \arcsec~(5 pixel) section of the slit centered
upon each knot.  For the knots with sufficient signal to noise, the
observed line profiles were deconvolved with the instrumental line profiles 
that possessed a full width at half maximum (FWHM) of 12.7 km s$^{-1}$
as determined from the arc lamp lines.
Because the deconvolution assumed that the line profiles can be
approximated by gaussian profiles, and because asymmetries are clearly
apparent in some of the observed profiles, one should use the velocity
dispersion measures only as an indicator of the velocity spread in
individual knots.  In HH 38 and HH 46CjetA, high and low velocity
components are clearly resolved.  For HH 120A and HH 120B, the observed
asymmetric profiles were deconvolved assuming the presence of two gaussians at
each position.  Although fits of comparable goodness were possible by using
three or more gaussians as is the case with virtually any assymetric
profile, within the uncertainties of the fits the two-gaussian fits
represent the simplest possible analysis. 
Table 1 lists the heliocentric radial velocities for each
knot, and the FWHM of the emission lines associated with the knots.
Position-velocity plots for each of the slit positions are shown
as contour diagrams in Figs. 1-5. 
In Fig. 5, one can note
the position-velocity structure associated with knot A that appears
to consist of a high velocity (-39.4 km s$^{-1}$) feature and a weak
low velocity (-5.0 km s$^{-1}$) feature that is offset spatially by
 about 0.5\arcsec~
toward knot B.  Knot B exhibits an asymmetry in the sense of having
an extended blue wing.  The two-gaussian deconvolution of knot A yielded
a flux ratio of 2.0 for the -39.4 km s$^{-1}$ feature compared with the
-5 km s$^{-1}$ feature, and for knot B the ratio was 3.0 for the
4.3 km s$^{-1}$ feature compared with the -13 km s$^{-1}$ feature.

\section{Discussion}
\subsection{HH 38}

Surprisingly little work has been done on HH 38 since its discovery by
Haro (see \citep{her74}).  \citet{coh83} associated it with
the flow responsible for HH 43 upon the basis of an approximate
alignment of HH 43 and HH 38 with the putative exciting star (HH 43 IRS-1) that 
was discovered with infrared mapping to be located
about 1\arcmin~to the northwest of HH 43.  \citet{gre94} later
discovered that the putative exciting star is a binary.  More recently,
\citet{sta00} have reported infrared H$_{2}$ imaging of the region that reveals
a long string of HH objects (including HH 38, HH 43, and HH 64) aligned 
with a recently discovered far-IR/mm source (HH 43 MMS1) that is almost
certainly the exciting star for the system.  The new source is located 
about 3\arcmin~NW of HH 43 IRS-1, and shows the signatures of a class 0
protostellar source.  The \citet{sta00} image reveals an independent
flow from HH 43 IRS-1 that produces a faint bow shock NW of the star, but
which is not in the alignment of the HH 38, 43, and 64 system. 
\citet{eis97} 
present an optical [S II] image of the HH 38/43 region that 
reveals multiple knot structure in HH 38 that appears to correlate
with the H$_{2}$ structure seen in \citet{sta00}, but with the
northernmost structure (our knots A and B) appearing relatively fainter
than in H$_{2}$. 
The IR image of \citet{sta00}, taken in
1998, shows knot B to be somewhat fainter than in our image, and
a knot is also seen immediately to the south of our knot C which is
outside the restricted field of our direct image.
  Optical spectra of HH 38 obtained by Dopita,
Schwartz, \&  Cohen (1981 unpublished) show that the optical knots are
of low excitation with the [S II] emission intensity rivaling that of
H$\alpha$, and [N I] $\lambda$5200 emission appears 
comparable to the H$\beta$ emission intensity.
The assumed distance to the HH 38/43 complex is 450 pc \citep{sta00}.

The position-velocity diagrams for three knots in HH 38 (Fig. 1)
reveal double-line features expected in bow shocks.
\citet{dav00} find similar features in HH 7 and HH 33.  In the
case of HH 38, the weak feature at higher negative velocity is most likely 
identified with emission closer to the apex of a bow shock that has been
swept nearly to the jet speed.  By contrast, emission from the wings
of the bow shock will be dominated by ambient gas that has undergone little
acceleration through the oblique bow wave.  Much of the ambient H$_{2}$ is
evidently dissociated at the bow shock apex, resulting in relatively weak
emission.  However, due to the much lower shock velocities in the oblique
wings of the bow shock, the post shock temperatures are insufficient to
dissociate a large fraction of the H$_{2}$, leading to stronger emission.
An argument against the identification of the high velocity peaks with
the apex of a bow shock could be made upon the basis of the small
velocity dispersions in those features.  Divergence of post shock flow
at the apex should yield considerably broader lines than in the wings
of the bow shock.  We note, however, that there is considerable 
uncertainty associated with the line width measurements of the weak
high velocity features.
The velocity measurements in Table 1 indicate that the separations of the two
velocity components are about 45 km s$^{-1}$, 47 km s$^{-1}$, and 54 km s$^{-1}$
for knots A, B, and C, respectively.  This suggests that the shock velocity
at the apex may be $\geq$ 50 km s$^{-1}$.

The negative velocities seen in HH 38
are consistent with an object that has been ejected toward the foreground of
a dark cloud that harbors the exciting star.  It is curious, however, that
HH 43, situated between the exciting star and HH 38, shows small 
positive heliocentric velocities (3 km s$^{-1}$ $\leq$ v $\leq$ 24 km s$^{-1}$),
suggestive of motion predominantly in the plane of the sky
\citep{sch99}.  \citet{sta00} suggest that the motion of the HH objects
in this system is primarily in the plane of the sky, as evidenced by
the absence of a detectable bipolar CO flow.  It seems unlikely that
the precessional motion suggested for the components of HH 43 by
\citet{sch99} could give rise to the velocity discrepancy between HH 38
and HH 43.  It would be of considerable
interest to obtain optical emission line velocities for all of the HH objects
in this flow, and to confirm their expected motions away from the source
HH 43 MMS1 with proper motion measurements. 

\subsection{HH 46-47}

The HH 46/47 system, discovered by \citet{sch77}, is the prototype
system for isolated star formation in a Bok Globule.  The globule (ESO 210-6A)
harbors an embedded young binary system of 0\farcs26 separation and
component
luminosities of about $5~L_{\sun}$ and $7~L_{\sun}$ \citep{rei00}.
The distance to this system has been estimated variously at 350 pc
\citep{eis94} to 450 pc \citep{gra89}.
A bipolar jet
emanates from one of the components with the blueshifted visible jet flow to the northeast 
in the foreground of the globule.  Infrared observations reveal the redshifted
counterflow
penetrating through the globule to its backside where bow shocks associated with the
flow can be seen optically as they emerge in projection from behind the globule.  The
system has been studied in exquisite detail with {\it HST} imaging of the
visible flow
\citep{hea96}, and infrared H$_{2}$ imaging of the entire system \citep{eis94}.
Our study focuses upon three discrete components of the flows: HH 47A, a bright
bow shock at the terminus of the blueshifted
optical jet; HH 46Cjet, the redshifted counter jet proximate
to the exciting star; and HH 46F, located at the boundary of a cavity
formed by the redshifted flow inside the globule.

\subsubsection{HH 47A}

The first detection of H$_{2}$ emission in HH 47A was by \citet{sch83} 
with the identification of six fluorescent ultraviolet emission lines
produced by
Ly $\alpha$ pumping of H$_{2}$ molecules which in turn must first be
warmed to the second vibrational state of the ground electronic state.
The B-X (1,3), (1,6), (1,7), and (1,8) transitions were observed in the
wavelength range $\lambda\lambda$ 1250-1565~\AA.  The expected near-infrared
(NIR) H$_{2}$ quadrupole emission in HH 47A
 was first detected by \citet{wil90}, and
the NIR H$_{2}$ emission was mapped out in detail for the HH 46/47
system by \citet{eis94}.  \citet{cur95} employed {\it HST}
ultraviolet imaging and spectroscopy to investigate the emission from
HH 47A, confirming the presence of the B-X lines produced by Ly $\alpha$
fluorescence, and showing that a dominant portion of the continuum
UV emission is due to two-photon emission from neutral hydrogen.
\citet{hea96} have presented a detailed analysis of optical
{\it HST} images of the HH 47 flow, and \citet{har99} have
focused upon the optical structure and the UV/optical spectra of HH 47A.
\citet{mic98} have reported proper motion measurements of H$_{2}$
features in the system.  The H$_{2}$ knots for which there are visible
counterparts show approximately the same proper motions as the visible
knots as measured by \citet{eim94}.

An unanswered question in previous studies is that of the relation
of the optical emission seen in atomic lines to that of the NIR H$_{2}$
emission in HH 47A.  The registration of H$_{2}$ and [S II] images by
\citet{eis94} shows that the peak H$_{2}$ emission is located
at the bow shock, but is displaced about 2\arcsec~to the west of the
apex of the bow shock along the northern wing of the shock.  The image
reveals H$_{2}$ emission outlining faintly the opposite wing of the
bow shock to the southeast, with very faint emission elsewhere over the
HH 47A complex. 
The high spectral resolution H$\alpha$ measurements of HH 47A
reported by \citet{har90} were obtained with a slit in
position angle 57$\arcdeg$ aligned along the jet, whereas our 90$\arcdeg$
slit orientation captured only the peak H$_{2}$ emission along the
northern bowshock wing.  One of the two parallel slit positions used by
\citet{har90} appears to pass through the approximate position
of the peak H$_{2}$ emission.  The H$\alpha$ heliocentric velocity at
the H$_{2}$
peak position is seen to be about -79 km s$^{-1}$ in the position-velocity
diagram of \citet{har90}.  This agrees closely with
the velocity of -76.5 km s$^{-1}$ determined from our H$_{2}$ measurements, and 
suggests that the H$_{2}$ emission indeed arises from postshock gas.
A second slit position
used by \citet{har90} was offset 4\arcsec~to the south, passing
through the region of the Mach disk and closer to the apex of the bow
shock where the H$\alpha$ emission is stronger, and where the heliocentric
velocity is centered near -105 km s$^{-1}$.  The velocity structure
revealed by the optical data are fully consistent with the bow shock
model for HH 47A developed in more detail by \citet{har99}. 

The optical data differ from the infrared data by showing a larger
velocity dispersion ($\sim$60 km s$^{-1}$ FWHM estimated for H$\alpha$)
than the H$_{2}$ emission (19 km s$^{-1}$ FWHM).
It is possible that the atomic emission is weighted more
by brighter H$\alpha$ emission in the region of the apex where the
velocity dispersion would be expected to be greater due to postshock
flow divergence. Also, atomic emission would suffer additional thermal
broadening and possibly broadening due to charge exchange. 
The ideal case of the working surface of a jet involves jet gas which
passes through the Mach disk and then is directed backward along the side of
the jet in a shocked cocoon.  The shocked cocoon is separated from the
bow shock gas by a cooling layer that prevents mixing of the bow shock
and Mach disk gases.  Numerical calculations, however, have shown that
the cooling layer is subject to instabilities, and the Mach disk and
bow shock gases can undergo complex mixing as the cooling layer experiences 
fragmentation.  The fragmentation of the cooling layer in 
 HH 47A is clearly indicated in the [S II] and H$\alpha$
{\it HST} images of \citet{har99}.  The continuum ultraviolet
{\it HST} images of HH 47A \citep{cur95} show a close
morphological correspondence to the optical images.  This is not
surprising since the ultraviolet emission is dominated by two-photon
emission from neutral hydrogen that should trace the same emission
pattern as H$\alpha$.  The H$_{2}$ ultraviolet fluorescence could arise
from molecules mixed with the atomic gas, or from the region of
peak NIR H$_{2}$ emission if sufficient Ly $\alpha$ flux reaches the
northern wing of the bow shock.

The location of the peak NIR H$_{2}$ emission and the kinematics of
the H$_{2}$ gas support the contention of \citet{mor94} that
the molecular emission in HH 47A is caused by the presence of H$_{2}$
molecules in the wake of HH 47D into which the HH 47A jet is
propagating.  Evidently these molecules either survived the passage
of the HH 47D shock, or were produced by reformation in the wake of
of HH 47D.
In either case, the molecules were swept to a velocity of
about 210 km s$^{-1}$ with respect to the exciting star if one adopts
the model of \citet{har99} where the angle of the flow with
respect to the plane of the sky is 34$\arcdeg$ \citep{eis97}.
Moreover, as pointed out by
\citet{mor94}, the preshock H$_{2}$ must experience shielding
from ionizing radiation of the Gum Nebula since both HH 47A and
HH 47D appear to have moved beyond the rim of the globule into the
ionizing field of the Gum Nebula.  The
structure of the H$_{2}$ emission further suggests that the gas in
the wake of HH 47D (the HH 47A preshock gas) is inhomogeneous as might
be expected if H$_{2}$ reformation occurred in denser post shock
fragments after passage of the HH 47D shock.

Finally, it can be noted in the position-velocity diagram for H$_{2}$ in
HH 47A (Fig. 2) that a velocity gradient occurs ranging from
-77 km s$^{-1}$
 at peak emission to about -94 km s$^{-1}$ 
3\arcsec~ east along the bow shock near the apex.  This agrees with the
H$\alpha$ velocity at the apex seen in \citet{har90}, and is 
expected of bow shock kinematics where the maximum acceleration of
material occurs at the apex of the bow shock, with smaller
acceleration along the oblique wings of the bow shock.  Relative to
the cloud which has v$_{helio}$ = 21.9 km s$^{-1}$ according to
\citet{kui87}, 
the radial velocity of the apex is about -132 km s$^{-1}$.
With the proper motion of 182 km s$^{-1}$ for the apex reported by
\citet{eim94}, we compute an inclination to a sky plane at the rest
velocity of the cloud of about 36\arcdeg.

\subsubsection{HH 46 Counterjet}

The H$_{2}$ image of the HH 46 counterflow seen in \citet{eis94}
suggests that the narrow counterjet emerges at the base of an
ovoid cavity which has been excavated in the globule 
by a stellar wind or previous
episodic flows from the exciting star.  In their proper motion
study of the counterjet, \citet{mic98} identify the jet as
HH 46Cjet, and we adopt their nomenclature for the identification
of knots in the jet.  
  The [S II] images seen both in
\citet{rei91} and in \citet{eis94}
show a similar morphology, but with
diffuse [S II] emission surrounding a larger portion of the
sharply defined counterjet.  \citet{rei91} have secured 
high dispersion optical spectra of this region of the
counterflow, finding a nearly constant velocity for the jet
(v$_{helio}$ $\sim$ +111 km s$^{-1}$), but a decelerating velocity
from about 111 km s$^{-1}$ near the base of the jet to about 50
km s$^{-1}$ 7\arcsec~downstream for the {\it diffuse} [S II] component.
It is suggested that the diffuse decelerating flow is due to the
interaction of a stellar wind with cavity walls. 

Our data (Fig. 3) exhibit a striking bifurcation of the H$_{2}$ velocity
between the ``base" of the jet and the jet itself, a feature that 
is seen in the low dispersion H$_{2}$ spectra of \citet{eis00}. 
The diffuse structure (HH46CjetA) that we refer to as the ``base"
of the jet exhibits
v$_{helio}$ = 31.6 km s$^{-1}$.
This is only about 10 km sec$^{-1}$ greater than
 the velocity of the parent cloud (and presumably
the velocity of the exciting star) found by \citet{kui87} to
be v$_{helio}$ = 21.9 km s$^{-1}$.  There is also faint H$_{2}$ emission
in this diffuse knot at the jet heliocentric velocity of about 
116 km s$^{-1}$.  Although it
is not clearly evident in the position-velocity contour diagram
(Fig. 3), in the original spectrum there is a hint of faint emission
connecting the low and high velocity features at the base of the jet.

In combination with the detailed high resolution optical spectra of the
HH46Cjet obtained by \citet{rei91} and the proper motion study of the
HH46Cjet by \citet{mic98}, our data are difficult to interpret in terms
of a coherent model.  First, we see no evidence for a diffuse
decelerating component along the jet as seen in [S II] by \citet{rei91}.
The [S II] deceleration is best seen in the \citet{rei91} spectrum which
was obtained with a slit position off the axis of the jet, but cutting
through the diffuse emission to the northwest of the jet.  Although it
is possible that our spectrum was not deep enough to uncover this faint
diffuse component, one can note in the H$_{2}$ and [S II] images of
\citet{eis94} that indeed the [S II] emission appears to surround a
larger portion of the jet than the H$_{2}$ emission that appears to
be more confined to the smaller diffuse structure of HH46CjetA. 
Because of bright overlying [S II] emission from the Gum Nebula at the
approximate velocity of the cloud, it is not possible to determine from
the \citet{rei91} spectra whether the ``base" of the jet exhibits a
[S II] feature which would correspond to the bright low velocity feature
which we see in H$_{2}$ (Fig. 3).  It is possible that the low velocity
H$_{2}$ component is due to highly oblique shocks in dense material
that is located at or near the orifice where the flow opens into
the ovoid cavity.  The low velocity, however, suggests that it is the
ambient cloud material that is being shocked, perhaps through prompt
entrainment, and accelerated to about 10 km sec$^{-1}$ relative to the
cloud.  The [Fe II] and H$_{2}$ spectra of the counterjet reported by
\citet{fer00} suggest that the jet base feature could be the
result of the most recent episodic ejection.  Fernandes finds strong
infrared [Fe II] emission in the feature at the velocity (v$_{helio}$
$\approx$ 100 km s$^{-1}$) of the jet.  Therefore the high velocity
[Fe II] emission could represent the head of a bow shock where most
of the H$_{2}$ is dissociated, with the lower velocity H$_{2}$ emission
arising from the wings of the bow shock where molecular material
is being entrained as the latest ejection of a confined jet opens into
the ovoid cavity.

A remaining mystery is that of the proper motion of 136 km s$^{-1}$
reported for the H$_{2}$ image of
 HH46CjetA by \citet{mic98}.  Our spectra indicate that
the flux from this feature 
is dominated by the low velocity H$_{2}$ component.  Therefore the
velocity and proper motion data do not yield a space velocity for the
feature which is commensurate with the jet itself.  Taken at face
value, the space velocity would indicate a flow inclination 
to the sky of only 4\arcdeg~ for knot A (relative to a plane in the
rest frame of the cloud).
Although the proper motions of the
jet knots B and D are uncertain, if we use the best proper motion
measurement which is for knot D (v = 229 km s$^{-1}$)
and a radial velocity of 92
km s$^{-1}$ (relative to the cloud velocity) from our measurements,
we find an inclination angle of about 22\arcdeg.  Within its large
uncertainty, this value agrees with the inclination found for the 
jet to the northeast of the exciting star.  That knot A probably
has the same inclination as the remainder of the jet is indicated
by the presence of the faint high velocity H$_{2}$ component and
the high velocity [Fe II] reported by Fernandes.

\subsubsection{HH 46F}

The object HH 46F lies along the southeastern flank of the ovoid cavity
swept out by the counterflow.  In the infrared images of
\citet{eis94} and \citet{eis00},
the object is seen to be clearly off-axis of the main flow.  One is
given the impression that the material in HH 46F may represent ambient
cloud material that has been entrained in the far flanks
of a bow shock, the apex of which has now propagated to the southwestern
tip of the ovid cavity about 1.25\arcmin~ from HH 46.  Another possibility
is that an inhomogenous stellar wind in the cavity is producing
entrainment of molecular material at the boundary of the cavity.
 The heliocentric velocity of the material measured from the H$_{2}$ emission
ranges from about 23 to 31 km s$^{-1}$ (see Fig. 4). 
This is slightly greater than the velocity of the
ambient cloud material as would be expected if cloud material has
been entrained by the {\it receding} outflow.   

\subsection{HH 120}

The second HH system that was discovered in a cometary globule in the
Gum Nebula is HH 120 in CG-30 \citep{rei81,pet84}.
The distance to this system is estimated to be 450 pc
\citep{gra89}.  In optical emission,
the object extends about 6\arcsec~with an east-west orientation, and
exhibits a protrusion to the north at the western end of the nebula.
A strong infrared source (IRS 4) discovered by \citet{pet84} is
generally believed to be the exciting source, and is located about
8.5\arcsec~to the ESE of the brighter western knot A in HH 120. 
The near
infrared images obtained by \citet{gra89} and \citet{gre94}
clearly indicate a luminous connection between IRS 4 and the HH nebula.
\citet{pet84} found that the spectrum of the eastern portion of the
nebula (his position B, 5\arcsec~east of position A) revealed evidence
for a red continuum with features typical of a T Tauri star.  His
infrared aperture photometry identified an infrared source (IRS 3) near
that position.  The subsequent imaging by \citet{gra89} and
\citet{gre94} suggests that the IRS 3 source may in fact be a reflection
nebula.  The polarimetry reported by \citet{sca90} confirms the
presence of a reflection nebula, and the geometry suggests that light
from IRS 4 is reflected along the walls of a rather narrow cavity that 
has presumably been formed by the flow responsible for HH 120.  The
situation appears somewhat analogous to the HH 46 reflection nebula.
Also, Pettersson found that HH 120A revealed a very low excitation
optical spectrum similar to that of HH 47A, furthering the analogy with
the HH 46/47 system.

Gredel's infrared H$_{2}$ image of HH 120 shows a morphology
essentially identical to that seen in Fig. 5 of this study.  Curiously,
although Gredel found significant infrared [Fe II] emission at the
location of HH 120A, there is no indication of significant [Fe II]
emission at the position of H$_{2}$ knot B (Fig. 5).  For clarification,
we note that the separation of knots A and B in the H$_{2}$ images is
about 2.8\arcsec, so the spectroscopic position B identified by
\citet{pet84} is actually centered about 2.2\arcsec~to the east
of H$_{2}$ knot B.  Gredel's infrared [Fe II] image shows the presence of
infrared [Fe II] knots along the channel between IRS 4 and knot A, a morphology
that is not evident in the optical image seen in \citet{pet84}.
We interpret this to mean that HH emission is occurring along the
outflow axis, but that the extinction is sufficiently high closer to
IRS 4 to obscure the optical components of the emission.

From his optical spectroscopy, Pettersson found a heliocentric
radial velocity of -42 km s$^{-1}$ $\pm$ 12 km s$^{-1}$ for knot A.
From the H$_{2}$ emission, we find a heliocentric velocity of -39
km s$^{-1}$ for knot A.  In addition, we note evidence in Fig. 7
for a secondary condensation at v$_{helio}$ = -5 km s$^{-1}$ displaced
about 0\farcs5 toward the flow source.  This would be consistent with
emission from a clump of material in a bow shock wing extending back
from the apex at position A.  
  The close agreement between the atomic and H$_{2}$ 
velocities at the apex indicates that, like HH 47A, molecular gas has passed
through the bow shock and accelerated to the same velocity as the
atomic gas.  The very low excitation character of both HH 47A and
HH 120A suggests that only partial dissociation of molecules occurs
near the apex of the bow shock.  It can be noted that the velocity
dispersion of the H$_{2}$ emission in HH 120A is somewhat greater
($\sim$30 km s$^{-1}$) than found in most of the other HH knots in this study.
This could be the result of excitation at the apex of a bow shock
where the velocity dispersion is expected to be a maximum, and the
value probably represents the approximate shock velocity of the gas.
  In the case of HH 120, there are no proper motion measurements which would
permit calculation of the angle of the flow with respect to the plane
of the sky.  The heliocentric velocity of the cloud is 22.3 km s$^{-1}$
as determined from CO measurements
\citep{pet84}, so the gas in HH 120A is moving at least 60 km s$^{-1}$
with respect to the presumed velocity of the exciting star.  Pettersson
finds little evidence for reddening in HH 120A, so it is possible that
the angle of ejection from IRS 4 is greater than 45\arcdeg~ with respect
to the plane of the sky, given the short projected distance between IRS 4
and HH 120A.
With the low excitation character of the
nebula, it is thus possible that the flow is moving into the wake of a
previous outflow.

The heliocentric velocity of the dominant component of
 knot B is about 4 km s$^{-1}$, again
in satisfactory agreement with that found for the optical lines by
Pettersson.  As noted previously, however, Pettersson's measurement
was made about 2\arcsec~east of the H$_{2}$ knot B, and his value
of 10 km s$^{-1}$ $\pm$ 25 km s$^{-1}$ involved only a few weak lines
that were detected at that position.  Knot B is immediately adjacent
to the axis of the flow from IRS 4 to HH 120A, and therefore may
represent an ambient molecular clump that has been swept through
a bow shock wing extending to the east from HH 120A.  The clump has apparently
been excited and accelerated to a radial velocity of about 18 km s$^{-1}$
with respect to the cloud.  The lack of a distinct optical emission knot
or a NIR [Fe II] knot at
the position of B indicates that the shock velocity is very low, and
that few of the molecules are dissociated in the shock.  This
is what one expects in the extended oblique wing of a bow shock.  If
this interpretation is correct, it highlights the importance of
the inhomogeneous distribution of material (clumps) in the preshock
medium, that in fact may have resulted from fragmentation in the
passage of earlier shocks.  The reflection found by Pettersson in
the vicinity of knot B may be related to the higher density material
implied by the H$_{2}$ knot.  The deep NIR H$_{2}$ image of the CG-30
globule obtained by \citet{hod95} shows the presence of additional groups
of emission knots that do not align with the IRS 4 - HH 120 system.
Proper motion and radial velocity data will be required to determine
the probable origin of these structures which suggest that there may
be multiple YSOs in CG-30.

\section{Summary}

The results of this study indicate that H$_{2}$ emission in HH
objects kinematically traces that seen in atomic emission.  In
many cases, HH objects are propagating into the wakes of previous
flows, and evidence suggests that the wakes of these flows have
resulted in fragmentation and formation of molecular material
that becomes the preshock medium for subsequent HH flows.  The deposition
of momentum and energy into the ambient medium by an episodic jet
is apparently complex with multiple shocks overtaking previously
shocked material.  Previous studies have shown that, in most cases,
jets appear to have sufficient momentum and energy to drive the
molecular outflows seen in CO emission (see \citet{che95} for models
of jet-driven molecular flows).  There is evidence that
the coupling between the jet and the molecular medium is accomplished
both by prompt entrainment in bow shocks \citep{dav97}, and by
turbulent entrainment along the jet \citep{mic00}.

The double velocity peaks seen in HH 38 indicate that the sharp knots
probably consist of unresolved bow shocks, with the strongest H$_{2}$
emission from the lower velocity bow shock wings.  In HH 47A, the emission
from a region adjacent to the apex of the bow shock is resolved, and the
kinematics of the material appear to be fully consistent with that seen
in H$\alpha$ emission which represents prompt entrainment in the HH 47A
bow shock.  The HH 46Cjet counterflow shows a well-collimated, high
velocity beam (knots B-D), separated from a diffuse base (knot A) that
evidently represents a more recent ejection.  HH 46CjetA exhibits a strong
low velocity component indicative of excitation of ambient cloud material
in the oblique wings of a bow shock, with a high velocity component
representing the apex of the bow shock. 
  HH 46F can be interpreted either as entrainment
of ambient cloud material in the far wings of an extended bow shock
outlined by the ovoid cavity, or by entrainment of ambient gas by a stellar
wind which flows into the cavity.  Finally, HH 120 shows some features
which are similar to the HH 47 system, namely, as a flow that has
exited the front side of a globule and that is producing very low
excitation shocks due to propagation into the wake of a previous outflow.
Proper motion studies of both HH 38 and HH 120 will be
required to further illuminate the kinematics of these objects.

This work was carried out with support from the National Science
Foundation through grant AST 9417209.  We are indebted to the IRTF
staff for providing observing support and
trouble-free equipment, and to the IRAF help
desk at the NOAO for assistance in the data reduction.  Finally,
we thank the referee for insightful comments and suggestions for
improvements in presentation of the data.

\clearpage

\figcaption{Restricted direct H$_{2}$ image of HH 38 and the position~-
velocity (p-v) diagrams for HH 38.  The slit positions are indicated on the
direct image and labeled according to position in the p-v diagrams.
The orientation of the direct image and the direction to the exciting
star are indicated next to the direct image.  For knot A, the lowest
contour is at a flux level of 1.0$\times$10$^{-5}$ erg s$^{-1}$ cm$^{-2}$ sr$^{-1}$
\AA$^{-1}$, with a linear contour interval of 1.2$\times$10$^{-5}$ erg
s$^{-1}$ cm$^{-2}$ sr$^{-1}$ \AA$^{-1}$.  For knot B, the lowest contour is at
flux level 8.7$\times$10$^{-6}$ erg s$^{-1}$ cm$^{-2}$ sr$^{-1}$ \AA$^{-1}$, with
a contour interval of 2.3$\times$10$^{-5}$ erg s$^{-1}$ cm$^{-2}$ sr$^{-1}$ \AA$^{-1}$.
For knot C, the lowest contour is at the same flux level as knot B, but with a
contour interval of 2.6$\times$10$^{-5}$ erg s$^{-1}$ cm$^{-2}$ sr$^{-1}$ \AA$^{-1}$.
\label{fig1}}

\figcaption{Restricted direct H$_{2}$ image and p-v diagram for HH 47A.
Notations as in Fig. 1.  The flux level of the lowest contour is
1.4$\times$10$^{-5}$ erg s$^{-1}$ cm$^{-2}$ sr$^{-1}$ \AA$^{-1}$,
and the linear contour interval is 1.1$\times$10$^{-5}$ erg s$^{-1}$
cm$^{-2}$ sr$^{-1}$ \AA$^{-1}$. \label{fig2}}

\figcaption{Restricted direct H$_{2}$ image and p-v diagram for HH 46Cjet.
Notations as in Fig. 1.  The flux level of the lowest contour is
1.1$\times$10$^{-5}$ erg s$^{-1}$ cm$^{-2}$ sr$^{-1}$ \AA$^{-1}$,
and the linear contour interval is 1.0$\times$10$^{-5}$ erg s$^{-1}$
cm$^{-2}$ sr$^{-1}$ \AA$^{-1}$.  \label{fig3}} 

\figcaption{Restricted direct H$_{2}$ image and p-v diagram for HH 46F.
Notations as in Fig. 1.  The flux level of the lowest contour is
1.2$\times$10$^{-5}$ erg s$^{-1}$ cm$^{-2}$ sr$^{-1}$ \AA$^{-1}$,
and the linear contour interval is 1.0$\times$10$^{-5}$ erg s$^{-1}$
cm$^{-2}$ sr$^{-1}$ \AA$^{-1}$.  \label{fig4}} 

\figcaption{Restricted direct H$_{2}$ image and p-v diagram for HH 120.
Notations as in Fig. 1.  IRS 4 is the exciting source for HH 120.  The
flux level of the lowest contour is 1.2$\times$10$^{-5}$ erg s$^{-1}$
cm$^{-2}$ sr$^{-1}$ \AA$^{-1}$, and the linear contour interval is
3.4$\times$10$^{-5}$ erg s$^{-1}$ cm$^{-2}$ sr$^{-1}$ \AA$^{-1}$.
\label{fig5}}

\clearpage

\end{document}